\documentclass[aps,pre,preprint,groupedaddress]{revtex4-2}

\usepackage{graphicx}
\usepackage{bm}
\usepackage{hyperref}
\usepackage{physics}
\usepackage{amsfonts}
\usepackage{amsmath}
\usepackage{empheq}
\usepackage{amsthm}
\usepackage[english]{babel}
\usepackage{float}
\usepackage{lipsum} 
\usepackage{pgfplots}
\usepackage{relsize}

\pgfplotsset{compat=1.18}
\usepackage{tikz}
\usepackage{tikz-3dplot}
\usetikzlibrary{arrows.meta,calc,shadows.blur,positioning,angles,quotes}

\usetikzlibrary{decorations.pathreplacing}

\newtheorem{definition}{Definition}[section]

\newtheorem{remark}{Remark}
\theoremstyle{definition}

\begin{document}
\title{Free semigroup non-relativistic phase space states quantisation}

\author{Georgii Koniukov}
\email{kongosha@live.com, for correspondence or professional inquiries.}
\affiliation{}

\begin{abstract}
Quantisation is considered as an action of classical point operators on field states at the Planck scale of the phase space. 
\par Emphasis is placed on the sequence of physical variables. A "state language/Dirac notation" for classical mechanics is introduced—one that, in many cases, replaces the principle of stationary action for conservative systems. Observed classical and quantum dynamics are considered as different vectorisations of the free semigroup contribution of time. There is a dream of describing classical dynamics as complex transformations of state vectors over such a free semigroup, rather than solving differential equations. The transition from the alphabet to states, from states to state functions, and from state functions to measurable quantities is managed by projectors. An attempt is made to provide a derivation of the Legendre transformation. The trivial multiplication of the potential energy by the wave function in the coordinate representation and its non-trivial operator action in the momentum representation are obtained without a Fourier transform nor the Planck scale $\hbar$. The quantisation procedure is divided into two stages: first, the action of point-like classical operators on field functions; and second, accounting for the physically finite scale of the Planck length against the backdrop of the possible infinitesimality in the classical phase space. The concept of an angle between states is introduced, defining the measure of "quantumness". \par The free semigroup time contribution is studied in the context of the Born rule.
\end{abstract}
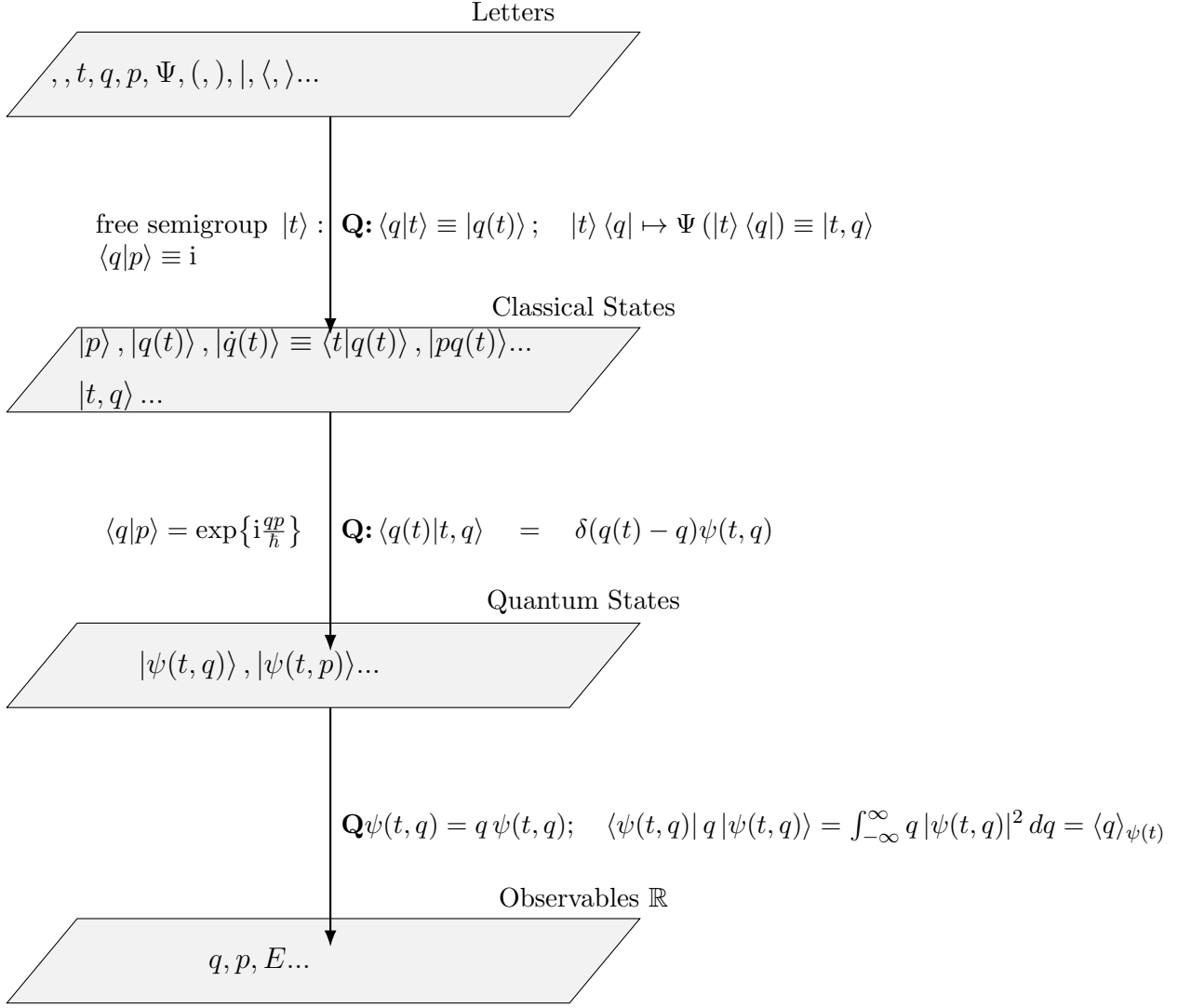
\maketitle

\tableofcontents

\section{Spaces}
\begin{tikzpicture}[
  >=Latex,
  plane/.style={draw, fill=gray!10},
  maparr/.style={->, thick},
  lab/.style={font=\small}
]

\begin{scope}[shift={(0,0)}]
  \draw[plane] (0,0) -- (8,0) -- (9,1.2) -- (1,1.2) -- cycle  node[midway, xshift=58pt] {$,,t,q,p,\Psi,(,),|,\langle,\rangle$...};
  \node[lab] at (7.2,1.50) {Letters};
\end{scope}

\begin{scope}[shift={(0,-4.2)}]
  \draw[plane] (0,0) -- (8,0) -- (9,1.2) -- (1,1.2) -- cycle
  node[midway, xshift=114pt, text width=7cm] {$\ket{p},\ket{q(t)},\ket{\dot{q}(t)}\equiv\braket{t}{q(t)},\ket{pq(t)}$...\\
  $\ket{t,q}...$};
  \node[lab] at (8.2,1.50) {Classical States};
\end{scope}

\begin{scope}[shift={(0,-8.4)}]
  \draw[plane] (0,0) -- (8,0) -- (9,1.2) -- (1,1.2) -- cycle
  node[midway, xshift=88pt] {$\ket{\psi(t,q)},\ket{\psi(t,p)}$...};
  \node[lab] at (8.2,1.50) {Quantum States};
\end{scope}

\begin{scope}[shift={(0,-12.6)}]
  \draw[plane] (0,0) -- (8,0) -- (9,1.2) -- (1,1.2) -- cycle
  node[midway, xshift=88pt] {$q,p,E$...};
  \node[lab] at (8.2,1.50) {Observables $\mathbb{R}$};
\end{scope}

\draw[maparr] (4.6,0.0) -- (4.6,-3.1) node[midway, right, lab] {$\textbf{Q:}\braket{q}{t}\equiv\ket{q(t)}; \quad \ket{t}\bra{q}\mapsto\Psi\left(\ket{t}\bra{q}\right)\equiv\ket{t,q}$};
 \node[lab] at (2.89,-1.55) {$\text{free semigroup}\ \ket{t}:$};
\node[lab] at (2,-2) {$\braket{q}{p}\equiv\mathrm{i}$};

\draw[maparr] (4.6,-4.2) -- (4.6,-7.6) node[midway, right, lab] {$\textbf{Q:}\bra{q(t)}\ket{t,q}\quad=\quad\delta(q(t)-q)\psi(t,q)$};
 \node[lab] at (2.8,-5.9) {$\braket{q}{p}=\exp{\mathrm{i}\frac{qp}{\hbar}}$};
\draw[maparr] (4.6,-8.4) -- (4.6,-11.8) node[midway, right, lab] { $\textbf{Q}\psi(t,q)=q\,\psi(t,q);\quad
    \bra{\psi(t,q)}
    q
    \ket{\psi(t,q)}
    =
    \int_{-\infty}^{\infty}
    q\,|\psi(t,q)|^2\,dq=\langle q\rangle_{\psi(t)}$}
    ;
\end{tikzpicture}
\section{Introduction}
Over the last 100 years, ideas of geometric quantization have been actively developing. In this introduction we mention only the most related to this work.
The Weyl–Wigner–Grönewold–Moyal phase-space formalism \cite{Moyal1949,Groenewold1946, Wigner1932, Weyl1927} that opened the connection between phase space and quantum mechanics. The "non-flat" complex geometry quantisation is done by Berezin  \cite{Berezin1974}. Recently, the understanding of phase space quantisation was enriched by De Gosson with the concept of quantum blobs, where the geometric nature of Planck's constant is shown \cite{deGosson2012}. 
\\ This work aims to predefine quantisation using Dirac state notation \cite{Dirac1939} for the phase space. We consider different spatial structurisation of the free semigroup contribution of time which once again relates to the von Neumann's formulation of quantum mechanics and algebras.\\ We will not explicitly repeat here the $\hat{p}\equiv-\mathrm{i}\hbar\frac{\partial}{\partial q}$ translational sense, but we will obtain $\braket{q}{p}=A\exp{\mathrm{i}\frac{qp}{\hbar}}$. 
\section{States in Classical Mechanics}
\subsection{Probable free semigroup contribution of non-parametrised time}
The concepts of time and the passage of time are not yet very well defined. We will examine a probable small aspect of the concept of time—perhaps an approximation—namely, the following contribution:
\begin{definition}{Time state column $\ket{t}$ contribution of non-parametrised time:}\label{t_definition}
\begin{equation}
        \ket{t}=  \begin{pmatrix}
    t_1 \\
    t_2 \\
    . \\
    . \\
    . \\
    t_N
  \end{pmatrix}=\sum\limits^N_{n=1}\ket{t_n}
\end{equation}
\end{definition}
\paragraph{a free semigroup under vertical concatenation equiped with a Pythagorean length.}
\par No scalar multiplication on $\ket{t}$ is assumed. Although $\ket{t}$ is not an element of a vector space, its concatenation
operation obeys a Pythagorean law. Equivalently, the squared length, which we will denote as $\|.\|^2$, hoping that this will not cause confusion, is additive under concatenation:
\begin{equation}
\|\ket{t_i}+\ket{t_j}\|^2=\|\ket{t_i}\|^2+\|\ket{t_i}\|^2
\end{equation}
 The free semigroup elements are not intrinsically orthogonal; rather, concatenation places their copies into mutually orthogonal slots defining the "basis" $\mathcal{B}=\{\ket{t_n}\}_{n=1}^{\infty}$.\\
\par Then, according to the Pythagorean theorem, the greater the length $\|\ket{t}\|$, the more $\ket{t_i}$ it contains, and vice versa.
\subsection{Trajectory (point) states}
\begin{definition}{Position vector as a function of time.}
    \begin{equation}\label{def:position}
       \bra{q}\ket{t}\equiv\begin{pmatrix}
    q_1 & q_2 & ... & q_\infty 
  \end{pmatrix}\circ
  \begin{pmatrix}
    t_1 \\
    t_2 \\
    . \\
    . \\
    . \\
    t_\infty
  \end{pmatrix}
  \equiv
  \begin{pmatrix}
    \braket{q_1}{t_1}\oplus\braket{q_2}{t_2} \oplus ... \oplus \braket{q_\infty}{t_\infty}
  \end{pmatrix}\mapsto \ket{q(t)}
\end{equation}
\end{definition}
\begin{definition}
     Zero-length element \(\mathbf{0}\) of the semigroup.\\ The letter \(\mathbf{0}\) is neither an identity element nor an absorbing zero of the semigroup; it is a distinguished generator that contributes zero to the Pythagorean length $\|\mathbf{0}\|=0$ while remaining part of the word.
    \begin{equation}\label{def:position_with_0}
       \bra{q}\ket{t}\equiv\begin{pmatrix}
    q_1 & q_2 & ... & q_\infty 
  \end{pmatrix}\circ
  \begin{pmatrix}
    t_1 \\
    . \\
    . \\
    \mathbf{0} \\
    . \\
    t_\infty
  \end{pmatrix}
  \equiv
  \begin{pmatrix}
    \braket{q_1}{t_1}\oplus\braket{q_2}{t_2} \oplus ...\oplus\mathbf{0}\oplus... \oplus \braket{q_\infty}{t_\infty}
  \end{pmatrix}\mapsto \ket{q(t)}+\ket{q(\tau)}
\end{equation}
\end{definition}
Zero-length symbol splits the trajectory into two, allowing the conditions for the same system to be reset; indicates that an interaction has occurred.
\begin{remark}
    Position and another vector $\dot{q}$ could be defined at the same time for classical states:
\end{remark}

\begin{equation}\label{position_velocity_state}
       \bra{q \dot q}\ket{t}\equiv\begin{pmatrix}
    q_1 & q_2 & ... & q_\infty \\
    \dot q_1 & \dot q_2 & ... & \dot q_\infty
  \end{pmatrix}\circ
  \begin{pmatrix}
    t_1 \\
    t_2 \\
    . \\
    . \\
    . \\
    t_\infty
  \end{pmatrix}
  \equiv
  \begin{pmatrix}
    q_1(t_1)\oplus q_2(t_2)\oplus ... \oplus q_\infty(t_\infty) \\
    \dot q_1(t_1)\oplus \dot q_2(t_2)\oplus ... \oplus \dot q_\infty(t_\infty)
  \end{pmatrix}\mapsto \ket{q\dot q(t)}
\end{equation}
Where $q_i(t_i),\ \dot q_j(t_j)$ are some functions with finite support.\\ 
And, more generally, let  \(\mathbb{K}=\mathbb{R}\) or
\(\mathbb{C}\).  The linear combinations,
\[
\mathbb{K}[\ket{t}]
:=
\left\{
\sum \braket{...}{t}
\right\},
\]
is now the vector space generated by \(\ket{t}\). 
\subsection{Configuraton-velocity space}
\begin{equation}
    \ket{q(t)}\otimes\ket{\dot q(t)}=\ket{q(t)\dot{q}{(t)}}
\end{equation}
\begin{figure}[H]
\centering

\begin{tikzpicture}[
    scale=1.1,
    vec/.style={-Latex, thick},
    edge/.style={thick},
    hidden/.style={thick, dashed}
]

\coordinate (A) at (0,0);
\coordinate (B) at (4,1.2);

\coordinate (C) at (1.2,2.2);
\coordinate (D) at (2.6,-1.5);



\draw[vec] (D) -- (C) node[pos=0.3, right] {$\ket{\dot{q}(t)}$};;




\draw[vec] (D) -- (B) node[midway, below right] {$\ket{q(t)}$};


\end{tikzpicture}

\caption{Configuration-velocity state space.}
\label{fig:double_right_triangle}

\end{figure}
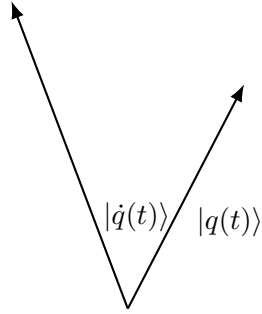

\begin{definition}{$\ket{t}-\text{entanglement reparametrisation:}$}
    \begin{equation}
    \ket{q(t)\dot{q}(t)}\longmapsto\ket{q\dot{q}(t)}
\end{equation}
\end{definition}

\subsubsection{Normalization}
Classical dynamics is defined by single state trajectory:
\begin{equation}
\bra{q(t)}\ket{q'(t')}\equiv\delta(q(t),q'(t'))
\end{equation}

\begin{equation}
    \ket{q(t)}\bra{q'(t)}= \delta(q(t),q'(t'))\circ\mathrm{id}
\end{equation}
\begin{definition}[Velocity]
\begin{equation}\label{def:velocity}
    \ket{\dot q (t)}\equiv\braket{t}{q(t)}
\end{equation}
\end{definition}
\begin{equation}\label{velocity_normalization}
   \braket{\dot{q}(t)}{\dot q (t)}\equiv\abs{C}^2
\end{equation}
And the correlation between velocity and position states as the dynamics of a given system:
\begin{equation}
    \braket{q(t)}{\dot{q}(t)}=C; \quad C\in\mathbb{C}, \  CC^*=\abs{C}^2 \in \mathbb{R_+}
\end{equation}
\begin{equation}\label{tangent_realtion}
    \ket{\dot q(t)}=\ket{q(t)}\braket{q(t)}{\dot q (t)}=C\ket{q(t)} 
\end{equation}
Next we define "free" differentials that have relation to the partial non-commutative derivative in free probability theory \cite{Mingo2017}.
\begin{definition}["Free" differentials/Conjugates over the quaternions $\mathbb{H}$]
\begin{equation}
\boxed{
    i\cdot\bra{t}, \ j\cdot\bra{\dot{q}},  \ k\cdot\bra{q}
    }
\end{equation}
And
\begin{equation}
    \braket{q}{q}=\braket{\dot{q}}{\dot{q}}=1
\end{equation}
Note, that without the semigroup $\ket{t}$, the $\ket{\dot{q}}$ is not the trajectory velocity state $\ket{\dot{q}(t)}$.\\
\subsubsection{"Free" Euler-Lagrange equation}
If we apply:
\begin{equation}
\left(i\cdot\bra{t}\right)\left(\bra{1}\otimes j\cdot\bra{\dot{q}}\right)\ket{q\dot{q}(t)}=i\cdot j\cdot\braket{t}{q(t)}=\left(i\cdot j\right)\ket{\dot{q}(t)}=k\ket{\dot{q}(t)}
\end{equation}
On the other side:
$$
k\cdot\bra{{q}\otimes1}\ket{q\dot{q}(t)}=k\ket{\dot{q}(t)}
$$
These two equations yield a consequence for a given $\ket{q\dot{q}(t)}$:
\begin{equation}
     i\cdot\bra{t}\otimes \ j\cdot\bra{\dot{q}}=  \ k\cdot\bra{q}\otimes\bra{1} \quad \big(=k\cdot\bra{1}\otimes\bra{q}\big)
\end{equation}
\begin{equation}\label{conjugates_equation}
\boxed{
    \bra{t}\otimes\bra{\dot{q}}= \bra{q}\otimes\bra{1} =\bra{1}\otimes\bra{q}
     }
\end{equation}
is "free" Euler-Lagrange equation.\\
From now on, we shall not explicitly show the quaternion factors. However, they can always be restored for quantisation.
\end{definition}
\subsubsection{Augmentation}
We define the augmentation $\varepsilon$ $\forall q$:
\begin{equation}\label{def:augmentation}
\varepsilon|q\equiv q|\varepsilon\equiv q|q\quad\Rightarrow\quad\braket{\varepsilon}{q}\equiv\braket{q}{q}=1
\end{equation}
\begin{equation}
    \bra{\varepsilon}\left(
        \sum_j \alpha_j\ket{q_j}
    \right)
    =
    \sum_j\alpha_j.
\end{equation}
And the boundary $\partial_1$:
\begin{equation}\label{def:boundary}
    \partial_1\equiv
    \bra{\varepsilon}\otimes 1
    -
    1\otimes\bra{\varepsilon}
    \end{equation}
Its action, for example:
\begin{equation}
    \partial_1(\ket{q}\otimes\ket{\dot{q}})
    =\ket{\dot{q}}-\ket{q}.
\end{equation}

\subsubsection{Euler–Lagrange equation}
By using the convention of eq. \ref{position_velocity_state}, we emphasize here that $q(t)$ and $\dot{q}(t)$ belong to the same $\ket{t}$: $\ket{t}-$entangled or reparametrised:
\begin{equation}
    \ket{q(t)\dot{q}(t)}\mapsto\ket{q\dot{q}(t)}
\end{equation}
Dirac state notation can replace action principle for the conservative systems. Indeed, let's consider the state $\ket{S}$. Since in classical mechanics we consider $\ket{q\dot{q}(t)}$ that defines the whole Hilbert space and dynamics of the system in the state $\ket{S}$ we can define it as

\begin{definition}[Classical state]\label{def:classical_state}
\begin{equation}
\boxed{
\ket{q \dot{q}(t)}\braket{q \dot{q}(t)}
{S}}
\end{equation}
Lagrangian:
\begin{equation}
\boxed{
    L(q(t),\dot q(t))\equiv \braket{q \dot{q}(t)}
{S}}
\end{equation}
And "free differential projector" $\varepsilon\dd\varepsilon$:
\begin{equation}
\bra{t\dot{q}}\varepsilon\dd\varepsilon\!\ket{S}\equiv\frac{\partial}{\partial t\big|_{\ket{S}}}\circ\frac{\partial}{\partial\dot{q}\big|_{\ket{S}}}
\end{equation}
\begin{equation}\label{state_differential}
    \boxed{
\varepsilon\dd\varepsilon\!\ket{q \dot{q}(t)}\braket{q \dot{q}(t)}
{S}}
\end{equation}
\end{definition}

Let's condider now how "free" Euler-Lagrange equation \ref{conjugates_equation} acts with the free differential projector on the state \ref{state_differential}:
\begin{equation}
    \bra{t\dot{q}}\varepsilon\dd\varepsilon\!\ket{q \dot{q}(t)}\braket{q \dot{q}(t)}
{S}\equiv\frac{\partial}{\partial t\big|_{\ket{q \dot{q}(t)}}}\circ\frac{\partial}{\partial\dot{q}\big|_{\ket{q \dot{q}(t)}}}L(q(t),\dot{q}(t))=\frac{d}{dt}\frac{\partial}{\partial\dot q}L(q(t),\dot{q}(t)
\end{equation}
And
\begin{equation}
     \bra{q\otimes1}\varepsilon\dd\varepsilon\!\ket{S}\equiv\frac{\partial}{\partial q\big|_{\ket{S}}}
\end{equation}
\begin{equation}
       \bra{q\otimes1}\varepsilon\dd\varepsilon\!\ket{q \dot{q}(t)}\braket{q \dot{q}(t)}
{S}\equiv\frac{\partial}{\partial q\big|_{\ket{q \dot{q}(t)}}}L(q(t),\dot{q}(t))=\frac{\partial}{\partial q}L(q(t),\dot{q}(t))
\end{equation}
Thus, for the given state we got the functions equality:
\begin{equation}
    \frac{d}{dt}\frac{\partial}{\partial\dot q}L(q(t),\dot{q}(t))=\frac{\partial}{\partial q}L(q(t),\dot{q}(t))
\end{equation}
\begin{remark}
   Lagrangian is often defined at the boundary (eq. \ref{def:boundary}):
   \begin{equation}\label{rem:boundary}
       \bra{S}\partial_1\ket{q\dot{q}}=K(\dot{q})-U(q)\equiv L(q,\dot{q})
   \end{equation}
\end{remark}
\subsubsection{Potential and Kinetic energies on the configuration-velocity space}
\begin{definition}[Potential energy]
\begin{equation}
     \hat{U}\equiv {\bra{U\otimes\dot{q}}}
\end{equation}
\begin{equation}
        {\bra{U\otimes\dot{q}}}\ket{q\dot{q}(t)}=U\ket{q(t)}=\ket{U\ket{q(t)}}=U(q(t))
\end{equation}
\end{definition}
\begin{definition}[Kinetic energy]
\begin{equation}
     \hat{K}\equiv \bra{q\otimes K}
\end{equation}
\begin{equation}
        \bra{q\otimes K}\ket{q\dot{q}(t)}=K\ket{\dot q(t)}=\ket{K\ket{\dot q(t)}}=K(\dot q (t))\equiv m\frac{\left(\dot{q}(t)\right)^2}{2}
\end{equation}
\end{definition}
\begin{remark}[The mass numbers the canonical variables, indicates their sequence.]
    \begin{equation}
        \frac{K(\dot q(t))}{U(q(t))}\propto m
    \end{equation}
    Because in $\ket{q(t)\dot q(t)}:\quad$ $\ket{q(t)}$ is the first, and $\ket{\dot q (t)}$ is the second.
\end{remark}

\subsection{Phase space}
\subsubsection{Momentum}
We want to avoid the time-derivative (tangent) relation between the $\ket{q(t)}$ and $\ket{\dot q (t)}$ in eq. \ref{tangent_realtion}, and to build the new space based on two more independent variables. To do this, we will seek a different variable $\ket{p}$ to project the the same state $\ket{S}$ (def. \ref{def:classical_state}), and this variable should be $\bra{t}$-independent to $\ket{q}$. In the states space, these two conditions give eq.\ref{p:first_condition} and eq.\ref{p:second_condition}:
\begin{equation}\label{p:first_condition}
\ket{p}{\bra{1\otimes\dot q}}\ket{q\dot q (t)}=\ket{pq(t)}
\end{equation}
Since we do not want to have the tangent relation between $\ket{p}$ and $\ket{q}$ we can choose and fix the angle between them; we can choose the same angle for any Lagrangian system/state $L(q,\dot{q})$ and define the convention:
\begin{equation}\label{p:second_condition}
    \braket{q}{p}\equiv \mathrm{i} \iff \ket{p}\perp\ket{q}
\end{equation}
We will define $\ket{p}$ as a function as well:
\begin{equation}
\ket{p}{\bra{1\otimes\dot q}}\varepsilon\dd\varepsilon\!\ket{q\dot q (t)}\braket{q\dot q (t)}{S}= \ket{p}\frac{\partial}{\partial \dot q}\ L(q(t),\dot{q}(t))=\ket{\ket{p}\frac{\partial L(q(t),\dot{q}(t))}{\partial \dot q}\ }
\end{equation}
\begin{equation}\label{def:momentum}
    \ket{\ket{p}\frac{\partial L(q(t),\dot{q}(t))}{\partial \dot q}\ }=\ket{p} \quad\text{only if}\quad p\equiv\frac{\partial L(q(t),\dot{q}(t))}{\partial \dot q}
\end{equation}
Here we defined the convention:
\begin{equation}\label{def:ket_convention}
    \ket{f(p)\ket{p}}\equiv f(\ket{p}), \quad\text{and}\quad \ket{f(q)\ket{p}}\equiv f(q).
\end{equation}
According to the possibility of the simultaneous point state existence (equation \ref{position_velocity_state}) we can assign the same $\ket{t}$ from $\ket{q(t)}$ to the row of $\ket{p}$, and we will get $\ket{p(t)}\equiv\braket{p}{t}$. \\
However, for now, to study the geometrical meaning of $p$ we will use its $\bra{t}-$independence and choose $\ket{t}\mapsto\ket{t_p}$ to conserve the orthogonality $\ket{p(t_p)}\perp\ket{q(t)}$, thus
\begin{equation}
\braket{q}{p}=\mathrm{i} \Rightarrow \braket{q(t)}{p(t_p)}=\mathrm{i}r,\ \text{for some}\ r\in\mathbb{R}, \ \text{and} \ \braket{p(t_p)}{p(t_p)}=r^2    
\end{equation}
For example, the (flat) Kähler potential is calculated using Pythagoras’ theorem; and, after, the reparametrisation:
\begin{equation}
\braket{q(t)}{q(t)}+\braket{p(t_p)}{p(t_p)}\longmapsto \rho\equiv q(t)^2+p(t)^2
\end{equation}
Thus, the geometry is now simplified (see eq.\ref{symplectic_form}): 
\begin{equation}
    L(q,\dot{q})\mapsto\rho(q,p)=q(t)^2+p(t)^2
\end{equation}
\subsubsection{Legendre transformation}
In order define the Legendre transformation (Fig. \ref{fig:double_right_triangle}) we consider the boundary identity defining $\operatorname{Im}\partial_2$ on the state sequence $\ket{p}\to\ket{q}\to\ket{\dot{q}}$:
\begin{equation}\label{eq:topology_simplex}
\partial_2\ket{p\otimes q\otimes \dot{q}}
=\ket{p\otimes q}+\ket{q\otimes\dot{q}}-\ket{p\otimes \dot{q}} \in \operatorname{Im}\partial_2
\end{equation}
And
\begin{equation}
        \partial_1\ket{p\otimes\dot{q}}= \ket{\dot{q}}-\ket{p}=-(\ket{p}-\ket{\dot{q}})
\end{equation}
Thus,
\begin{equation}
    \partial_1\partial_2\ket{p\otimes q\otimes\dot{q}}=\ket{q}-\ket{p}+\ket{\dot{q}}-\ket{q}+\ket{p}-\ket{\dot{q}}+\ket{p}=0
\end{equation}
And then in
\begin{equation}
    \operatorname{\operatorname{H_1}}\equiv{\operatorname{Ker}\partial_1}/{\operatorname{Im}\partial_2}: \quad \ket{p\otimes q}+\ket{q\otimes\dot{q}}=\ket{p\otimes \dot{q}}
\end{equation}
is "free" Legendre transformation.\\
Then if we act on the state $\bra{S}$ in $\operatorname{\operatorname{H_1}}$:
\begin{align}
    \braket{S}{p\otimes q}+\braket{S}{q\otimes\dot{q}}=\braket{S}{p\otimes\dot{q}}
\end{align}
Note, that we initially had a sequence $\ket{pq\dot{q}},(p\to q\to\dot{q})$ where we have the factorisation over $\ket{q}$ in the middle, then $\braket{S}{pq}/(p\to q\to\dot{q})$ allows us to define for many systems final $\partial_1$-boundary potential $U(q)$ inside an arbitrary state function $H(p,q)$. $\braket{S}{q\dot{q}}/(p\to q\to\dot{q})$ allows us to define for many systems initial $\partial_1$-boundary potential $U(q)$ (eq. \ref{rem:boundary})  inside an arbitrary state function $L(q,\dot{q})$. However, for the $\braket{S}{p\dot{q}}/(p\to q\to\dot{q})$ it should be impossible to define such isolated boundary potentials $W(p),K(\dot{q})$ for any system, because they coincide with the full factorisation direction. This is why we will search $\braket{S}{p\dot{q}}$ in a form of $p\dot{q}F(p,\dot{q})$.
\begin{equation}
    H(p,q)+L(q,\dot{q})= p\dot{q}F(p,\dot{q})
\end{equation}
$\frac{\partial}{\partial \dot{q}}:$
\begin{equation}
    p=pF(p,\dot{q})+p\dot{q}\frac{\partial F}{\partial\dot{ q}}
\end{equation}
\begin{equation}
    1-\dot{q}\frac{\partial F}{\partial\dot{ q}}=F(p,\dot{q})
\end{equation}
Then
\begin{equation}
    F(p,\dot{q})=1+\frac{C(p)}{\dot{q}}
\end{equation}
Then 
\begin{equation}
    p\dot{q}F(p,\dot{q})=p\dot{q}+pC(p)
\end{equation}
where $pC(p)$ is an isolated (boundary) potential inside $\braket{S}{p\dot{q}}/(p\to q\to\dot{q})$, thus, $C(p)\equiv0$, and
\begin{equation}
        H(p,q)+L(q,\dot{q})= p\dot{q}
\end{equation}
\begin{remark}[Tensor rescaling property of momentum]
After introducing a new vector $\ket{p}$, we still describe the same state, and $\ket{p}$ has a rescaling meaning:
\begin{equation}
    \partial\ket{q\otimes\dot{q}}=\partial\big[\ket{p\otimes\dot{q}}-\ket{p\otimes q}\big]
\end{equation}
\begin{equation}
\partial\ket{q\otimes\dot{q}}=\partial\ket{p\otimes\partial\ket{q\otimes\dot{q}}}
\end{equation}
\begin{equation}
    \ket{q\dot{q}}\cong\ket{pq \ p\dot{q}}
\end{equation}
\end{remark}
\begin{remark}{$\ket{p(t_p)\otimes\dot{q}(t)}\perp\ket{q(t)\otimes\dot{q}(t)}$}
\begin{equation}
    \braket{p(t_p)\otimes\dot{q}(t)}{q(t)\otimes\dot{q}(t)}=\braket{p(t_p)}{q(t)}\braket{\dot{q}(t)}{\dot{q}(t)}=(-\mathrm{i}r)\cdot \abs{C}^2 \in \mathrm{i}\mathbb{R}
\end{equation}
\end{remark}
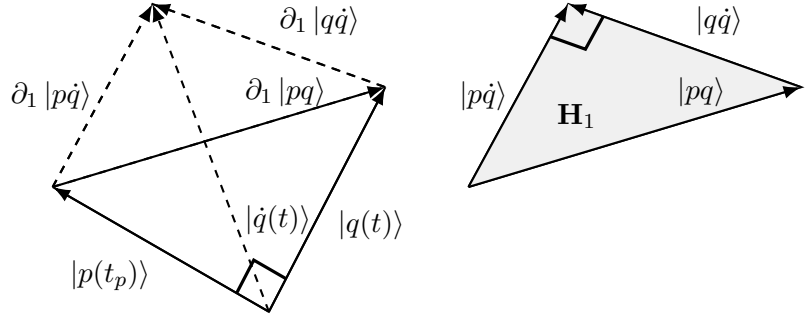
\begin{figure}[H]
\centering

\begin{tikzpicture}[
    scale=1.1,
    vec/.style={-Latex, thick},
    edge/.style={thick},
    hidden/.style={thick, dashed}
]

\coordinate (A) at (0,0);
\coordinate (B) at (4,1.2);

\coordinate (C) at (1.2,2.2);
\coordinate (D) at (2.6,-1.5);
\coordinate (E) at (5,0);
\coordinate (F) at (9,1.2);

\coordinate (G) at (6.2,2.2);

\draw[edge,hidden] (A) -- (C) -- (B);
\draw[edge,hidden] (A) -- (B);

\draw[edge] (A) -- (D) -- (B);
\draw[edge] (E) -- (F) -- (G) -- (E);
\fill[gray!12] (E) -- (F) -- (G) -- cycle;
\node at (barycentric cs:E=2,F=1,G=1) {$\textbf{H}_1$};
\draw[vec] (E) -- (G) node[midway, left] {$\ket{p\dot{q}}$};
\draw[vec] (F) -- (G) node[midway, above right] {$\ket{q\dot{q}}$};
\draw[vec] (E) -- (F) node[pos=0.7, above] {$\ket{pq}$};

\draw[vec,hidden] (D) -- (C) node[pos=0.3, right] {$\ket{\dot{q}(t)}$};

\pic [draw, line width=1.2pt, angle radius=14pt] {right angle = E--G--F};
\pic [draw, line width=1.2pt, angle radius=14pt] {right angle = A--D--B};

\draw[vec,hidden] (A) -- (C) node[midway, left] {$\partial_1\ket{p\dot{q}}$};
\draw[vec,hidden] (B) -- (C) node[midway, above right] {$\partial_1\ket{q\dot{q}}$};

\draw[vec] (D) -- (A) node[midway, below left] {$\ket{p(t_p)}$};
\draw[vec] (D) -- (B) node[midway, below right] {$\ket{q(t)}$};

\draw[vec] (A) -- (B) node[pos=0.7, above] {$\partial_1\ket{pq}$};


\end{tikzpicture}

\caption{Legendre transformation.}
\label{fig:double_right_triangle}

\end{figure}

\begin{remark}{Scalar multiplication of boundaries can be represented as scalar multiplication (with conjugation)}
\begin{equation}
\partial\ket{p\otimes q}\mapsto1-ir; \quad\partial\ket{q\otimes\dot{q}}\mapsto C-1;\quad\partial\ket{p\otimes\dot{q}}\mapsto C-ir; 
\end{equation}
\end{remark}

\subsubsection{Potential and Kinetic energies on the phase space}
We need to define the energy that acts on a classical state $\ket{pq(t)}$, thus, by using the convention of eq. \ref{position_velocity_state}, we emphasize here that $q(t)$ and $p(t)$ will now again belong to the same $\ket{t}$ explicitly; $\ket{t}-$entangled or reparametrised:
\begin{equation}
    \ket{p(t_p)q(t)}\mapsto\ket{pq(t)}
\end{equation}
\begin{definition}[Potential energy]
\begin{equation}
     \hat{U}\equiv U{\bra{p\otimes1}}
\end{equation}
\begin{equation}
       U{\bra{p\otimes1}}\ket{pq(t)}=U\ket{q(t)}=\ket{U\ket{q(t)}}=U(q(t))
\end{equation}
\end{definition}
\begin{definition}[Kinetic energy]
\begin{equation}
     \hat{K}\equiv K\bra{1\otimes q}
\end{equation}
\begin{equation}
        K\bra{1\otimes q}\ket{pq(t)}=K\ket{p(t)}=\ket{K\ket{p(t)}}=K(p (t))\equiv\frac{1}{m}\frac{(p(t))^2}{2}
\end{equation}
\end{definition}
\begin{remark}[The mass numbers the canonical variables, indicates their sequence.]
    \begin{equation}
        \frac{K(p(t))}{U(q(t))}\propto \frac{1}{m}
    \end{equation}
    Because in $\ket{p q(t)}:\quad$ $\ket{p}$ is the first, and $\ket{q (t)}$ is the second.
\end{remark}

\subsection{Field}
To define the trivial field we will change the order of $\bra{q}$ and $\ket{t}$ states, and unlike in eq. \ref{position_velocity_state} it is impossible now to define position and velocity at the same time, since an $n\times1$ column cannot be multiplied by a $2\times n$ matrix in this order:
\begin{equation}
       \ket{t}\bra{q}\equiv
  \begin{pmatrix}
    t_1 \\
    t_2 \\
    . \\
    . \\
    . \\
    t_\infty
  \end{pmatrix}\circ
  \begin{pmatrix}
    q_1 & q_2 & ... & q_\infty 
  \end{pmatrix}
  \equiv
  \begin{pmatrix}
    (t_1, q_1) & (t_1,q_2) & ... & (t_1,q_\infty) \\
    (t_2,q_1) & (t_2,q_2) & ... & (t_2,q_\infty) \\
    .\\
    .\\
    .\\
  \end{pmatrix}=\sum\limits_{i,j}^\infty\ket{t_i}\bra{q_j}
\end{equation}
We will now define the arbitrary field  $\Psi$. "Field" means that it will respect $\ket{t}$ and position $\bra{q}$ independently. Once these $\ket{t}$ and $\bra{q}$ bases are fixed, we make the standard identification of a bilinear form with its representing matrix and denote both by $\Psi$.

\begin{multline}
        \Psi\equiv  \begin{pmatrix}
    \psi(t_1, q_1) & \psi(t_1,q_2) & ... & \psi(t_1,q_\infty) \\
    \psi(t_2,q_1) & \psi(t_2,q_2) & ... & \psi(t_2,q_\infty) \\
    .\\
    .\\
    .\\
  \end{pmatrix}=\sum_{i,j}^\infty\ket{t_i}\bra{t_i}\Psi\ket{q_j}\bra{q_j}=\sum\limits_{i,j}^\infty\psi(t_i,q_j)\ket{t_i}\bra{q_j}
\end{multline}
\section{Quantisation}
To explore how the trajectory defined point-state operators $\hat{U}$ and $\hat{K}$ can act on any  $\ket{...}$ field substate, we project these field states on to a basis of trajectory states: 
\begin{equation}
    \int \ket{q(t)}\bra{q(t)}\dd q(t) \equiv \mathbb{P}_{q(t)}
\end{equation}
\begin{equation}
    \int \ket{p(t)}\bra{p(t)}\dd p(t) \equiv \mathbb{P}_{p(t)}
\end{equation}
Now let's define the notation for an arbitrary field:
\begin{equation}
\boxed{
\ket{t,q}\equiv\ket{q}\bra{q}\Psi\ket{t}=\ket{q}\psi(t,q)
}
\end{equation}
Thus, correlation between point state and field state:
\begin{equation}\label{pointxfield}
\boxed{
    \bra{q(t)}\ket{t,q}\equiv\delta(q(t)-q)\cdot\psi(t,q)=\text{point function}\times\text{field function}
    }
\end{equation}
In this way, we can visualize $\ket{t}\bra{t}$ not as flat, but as existing within a volume—we are, in a sense closing the circuit by coupling the point state with the field state.
\begin{equation}
    \bra{p(t)}\ket{t,q}\equiv\delta(p(t)-q)\cdot\psi(t,q)
\end{equation}
Note, that $\braket{p(t)}{q}=\delta(p(t)-q)$, since $\ket{p(t)}$ and $\ket{q}$ here are not $\ket{t}-$entangled and not related to the same state in general, and are now just different letters. 
\begin{definition}{Field states}
\begin{equation}
\ket{t,q\otimes1}\ \text{or}\ \ket{1\otimes t,q}
\end{equation}
\begin{equation}
\ket{t,p\otimes1}\ \text{or}\ \ket{1\otimes t,p}
\end{equation}
\end{definition}
In the next subsection we will use the convention of eq. \ref{def:ket_convention}:
\begin{equation}
    \ket{f(p)\ket{p}}\equiv f(\ket{p}), \quad\text{and}\quad \ket{f(q)\ket{p}}\equiv f(q).
\end{equation}
\subsection{Potential energy}\label{quantum_potential_energy}
\begin{multline}
    \hat{U}\ket{t,q}= \braket{p\otimes U}{\mathbb{P}_{q(t)}|1\otimes t,q}=\int\braket{p\otimes U}{1\otimes q(t)}\braket{1\otimes q(t)}{1\otimes t,q}\dd q(t)=\\
    \int\bra{p}\cdot U\ket{q(t)}\delta(q(t)-q)\psi(t,q) \dd q(t)=\bra{p} U\ket{q}\cdot\psi(t,q)=\bra{1\bra{p}} \ \ \ket{U\ket{q}}\cdot\psi(t,q)=1\cdot{U(q)}\cdot\psi(t,q)
\end{multline}
\begin{multline}
    \hat{U}\ket{t,p}\equiv \braket{p\otimes U}{\mathbb{P}_{q(t)}|1\otimes t,p}=\int\braket{p\otimes U}{1\otimes q(t)}\braket{1\otimes q(t)}{1\otimes t,p}\dd q(t)=\\
    \int\bra{p}\cdot U\ket{q(t)}\delta(q(t)-p)\psi(t,p) \dd q(t)=\bra{p} \ U\ket{p}  \psi(t,p)=\bra{p} \ \ket{U\ket{p}}  \psi(t,p)=\bra{p} \ U(p)  \psi(t,p)=\\ \braket{\bra{p}U(p)}{\psi(t,p)}=U\left(\bra{p}\right)\ket{\psi(t,p)}
\end{multline}
\subsection{Kinetic energy}\label{quantum_kinetic_energy}
\begin{multline}
    \hat{K}\ket{t,q}\equiv \braket{K\otimes q}{\mathbb{P}_{p(t)}|t,q\otimes 1}=\int\braket{K\otimes q}{p(t)\otimes 1}\braket{p(t)\otimes 1}{t,q\otimes1}\dd p(t)=\\
 \int\frac{(p(t))^2}{2m}\bra{q}\delta(p(t)-q)\psi(t,q)\dd p(t)=\frac{q^2}{2m}\bra{q}\psi(t,q)=\braket{\frac{q^2}{2m}\bra{q}}{\psi(t,q)}=\frac{\bra{q}^2}{2m}\ket{\psi(t,q)}
\end{multline}
\begin{multline}
    \hat{K}\ket{t,p}\equiv \braket{K\otimes q}{\mathbb{P}_{p(t)}|t,p\otimes 1}=\int\braket{K\otimes q}{p(t)\otimes 1}\braket{p(t)\otimes 1}{t,p\otimes1}\dd {p(t)}=\\
 \int\frac{(p(t))^2}{2m}\bra{q}\delta(p(t)-p)\psi(t,p)\dd p(t)=\frac{p^2}{2m}\psi(t,p)\bra{q}=\frac{p^2}{2m}\psi(t,p)\bra{1\bra{q}}=\frac{p^2}{2m}\psi(t,p)
\end{multline}
Here we have that $U(\bra{p})$ is an operator without using the Fourrier transform nor the Planck scale $\hbar$. 
\subsection{Solid angle between states}
In classical Hamiltonian mechanics, the dynamics is completely determined by the symplectic structure of phase space. The flat symplectic two-form is derived from the flat Kähler potential \cite{CannasDaSilva2001}:
\begin{equation}\label{symplectic_form}
\begin{aligned}
z_j&=q_j+\mathrm{i}p_j,
\qquad
\bar z_j=q_j-\mathrm{i}p_j,
\qquad
\rho=\sum_j |z_j|^2,
\\
\omega
&=\frac{\mathrm{i}}{2}\,\partial\bar\partial\rho
 =\frac{\mathrm{i}}{2}
   \sum_{j,k}
   \frac{\partial^2\rho}
        {\partial z_j\,\partial\bar z_k}\,
   dz_j\wedge d\bar z_k
 =\frac{\mathrm{i}}{2}\sum_j dz_j\wedge d\bar z_j
 =\sum_j dq_j\wedge dp_j .
\end{aligned}
\end{equation}
This encodes the canonical geometry, phase space integral $\int_\Sigma \omega$, and uniquely defines Hamiltonian vector fields (equations of motion).
This symplectic form can represent the infinitesimal oriented area element in phase space if calculated on infinitely small vectors. \\
However, because of quantum effects, it appears that this area cannot be infinitesimal, and we postulate here:
\begin{equation}
\boxed{
    \abs{\omega}_{\min}=\abs{dq \wedge dp}\approx\hbar
    }
\end{equation}
We mean that classical mechanics and, consequently, $\omega$ usually do not extend beyond this limit. 
\begin{figure}[H]
\centering
\begin{tikzpicture}

\pgfmathsetmacro{\R}{2}
\pgfmathsetmacro{\A}{2.2}
\pgfmathsetmacro{\B}{1.7}

\begin{axis}[
    view={138}{20},
    axis equal image,
    hide axis,
    scale=1.25,
    clip=false
]


\addplot3[
    surf,
    shader=flat,
    fill opacity=0.25,
    draw=gray!60,
    domain=-3:5,
    y domain=-2.5:4.5,
    samples=2,
    samples y=2
]
({x},{y},{0});

%

\addplot3[
    surf,
    shader=flat,
    fill=blue!15,
    opacity=0.22,
    draw=none,
    draw=blue!35,
    draw opacity=0.18,
    domain=0:360,
    y domain=0:180,
    samples=40,
    samples y=20
]
(
    {\R*sin(y)*cos(x)},
    {\R*sin(y)*sin(x)},
    {\R+\R*cos(y)}
);

%

\addplot3[
    fill=orange!40,
    fill opacity=0.45,
    draw=orange!80!black,
    very thick
]
coordinates {
    (0,0,0)
    (\A,0,0)
    (\A,\B,0)
    (0,\B,0)
    (0,0,0)
};

%
%
%
%
%

\addplot3[
    surf,
    shader=interp,
    fill=red!65,
    opacity=0.85,
    draw=red!50!black,
    mesh/rows=16,
    mesh/cols=16,
    domain=0:\A,
    y domain=0:\B,
    samples=16,
    samples y=16
]
(
    {\R*x/sqrt(\R^2+x^2+y^2)},
    {\R*y/sqrt(\R^2+x^2+y^2)},
    {\R-\R^2/sqrt(\R^2+x^2+y^2)}
);


\addplot3[
    solid,
    thin,
    black
]
coordinates {
    (-0.05,0,\R)
    (-0.05,0,0)
};
\addplot3[
    dashed,
    thick,
    red!70!black
]
coordinates {
    (0,0,\R)
    (0,0,0)
};

\addplot3[
    dashed,
    thick,
    red!70!black
]
coordinates {
    (0,0,\R)
    (\A,0,0)
};

\addplot3[
    dashed,
    thick,
    red!70!black
]
coordinates {
    (0,0,\R)
    (0,\B,0)
};

\addplot3[
    dashed,
    thick,
    red!70!black
]
coordinates {
    (0,0,\R)
    (\A,\B,0)
};


\draw[->, very thick]
    (axis cs:0,0,0)
    --
    (axis cs:4.5,0,0)
    node[anchor=north] {$\ket{q}$};

\draw[->, very thick]
    (axis cs:0,0,0)
    --
    (axis cs:0,4,0)
    node[anchor=north] {$\mathrm{i}\ket{p}$};


\addplot3[
    only marks,
    mark=*,
    mark size=1.7pt
]
coordinates {(0,0,0)};

\node[
    anchor=north east
]
at (axis cs:0,0,0) {$O$};

\node[
    anchor=west
]
at (axis cs:0,0,\R*0.5) {$\sqrt{\hbar}$};
\addplot3[
    only marks,
    mark=*,
    mark size=1.7pt
]
coordinates {(0,0,\R)};

\node[
    anchor=west,
    rotate=-10
]
at (axis cs:0,0,\R) {$\bra{t_p}$};
\node[
    anchor=south,
    rotate=5
]
at (axis cs:0.2,-0.5,\R-0.5) {$\ket{t_q}$};

\node[
    anchor=south
]
at (axis cs:{\A},0,0) {$\ket{q_k}$};

\node[
    anchor=west
]
at (axis cs:0,{0.75*\B},0) {$\mathrm{i}\ket{p_k}$};

\node[
    red!70!black,
    font=\large
]
at (
    axis cs:
    {\R*\A/sqrt(\R^2+(0.50*\A)^2+(0.50*\B)^2)},
    {\R*0.50*\B/sqrt(\R^2+(0.50*\A)^2+(0.50*\B)^2)},
    {\R-0.1*\R^2/sqrt(\R^2+(0.50*\A)^2+(0.50*\B)^2)}
)
{$\Omega$};

\end{axis}

\end{tikzpicture}
\caption{Quantum limit of the phase space states.}
\label{fig:quantum_limit}
\end{figure}
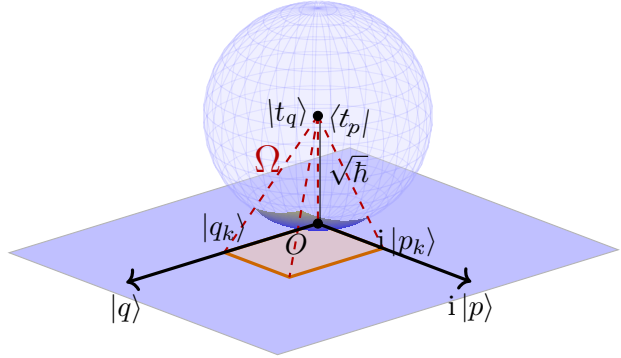


\begin{figure}[H]
\centering
\begin{tikzpicture}

\pgfmathsetmacro{\R}{2}
\pgfmathsetmacro{\A}{12.2}
\pgfmathsetmacro{\B}{11.7}

\begin{axis}[
    view={138}{20},
    axis equal image,
    hide axis,
    scale=1.25,
    clip=false
]


\addplot3[
    surf,
    shader=flat,
    fill opacity=0.25,
    draw=gray!60,
    domain=-3:12.5,
    y domain=-2.5:12,
    samples=2,
    samples y=2
]
({x},{y},{0});

%

\addplot3[
    surf,
    shader=flat,
    fill=blue!15,
    opacity=0.22,
    draw=none,
    draw=blue!35,
    draw opacity=0.18,
    domain=0:360,
    y domain=0:180,
    samples=40,
    samples y=20
]
(
    {\R*sin(y)*cos(x)},
    {\R*sin(y)*sin(x)},
    {\R+\R*cos(y)}
);

%

\addplot3[
    fill=orange!40,
    fill opacity=0.45,
    draw=orange!80!black,
    very thick
]
coordinates {
    (0,0,0)
    (\A,0,0)
    (\A,\B,0)
    (0,\B,0)
    (0,0,0)
};

%
%
%
%
%

\addplot3[
    surf,
    shader=interp,
    fill=red!65,
    opacity=0.85,
    draw=red!50!black,
    mesh/rows=16,
    mesh/cols=16,
    domain=0:\A,
    y domain=0:\B,
    samples=16,
    samples y=16
]
(
    {\R*x/sqrt(\R^2+x^2+y^2)},
    {\R*y/sqrt(\R^2+x^2+y^2)},
    {\R-\R^2/sqrt(\R^2+x^2+y^2)}
);


\addplot3[
    solid,
    thin,
    black
]
coordinates {
    (-0.05,0,\R)
    (-0.05,0,0)
};
\addplot3[
    dashed,
    thick,
    red!70!black
]
coordinates {
    (0,0,\R)
    (0,0,0)
};

\addplot3[
    dashed,
    thick,
    red!70!black
]
coordinates {
    (0,0,\R)
    (\A,0,0)
};

\addplot3[
    dashed,
    thick,
    red!70!black
]
coordinates {
    (0,0,\R)
    (0,\B,0)
};

\addplot3[
    dashed,
    thick,
    red!70!black
]
coordinates {
    (0,0,\R)
    (\A,\B,0)
};


\draw[->, very thick]
    (axis cs:0,0,0)
    --
    (axis cs:13.5,0,0)
    node[anchor=north] {$\ket{q}$};

\draw[->, very thick]
    (axis cs:0,0,0)
    --
    (axis cs:0,13.5,0)
    node[anchor=north] {$\mathrm{i}\ket{p}$};


\addplot3[
    only marks,
    mark=*,
    mark size=1.7pt
]
coordinates {(0,0,0)};

\node[
    anchor=north east
]
at (axis cs:0,0,0) {$O$};

\node[
    anchor=west
]
at (axis cs:0,-0.2,\R*0.3) {$\sqrt{\hbar}$};
\addplot3[
    only marks,
    mark=*,
    mark size=1.7pt
]
coordinates {(0,0,\R)};

\node[
    anchor=west,
    rotate=-7.5
]
at (axis cs:0,0,\R+0.1) {$\bra{t_p}$};
\node[
    anchor=south
]
at (axis cs:0.2,-1,\R-1.1) {$\ket{t_q}$};
\node[
    anchor=south
]
at (axis cs:{\A},0,0) {$\ket{q_0}$};

\node[
    anchor=west
]
at (axis cs:0,{0.75*\B},0) {$\mathrm{i}\ket{p_0}$};

\node[
    red!70!black,
    font=\large
]
at (
    axis cs:
    {\R*\A/sqrt(\R^2+(0.50*\A)^2+(0.50*\B)^2)},
    {\R*0.50*\B/sqrt(\R^2+(0.50*\A)^2+(0.50*\B)^2)},
    {\R-0.1*\R^2/sqrt(\R^2+(0.50*\A)^2+(0.50*\B)^2)}
)
{$\Omega$};

\end{axis}

\end{tikzpicture}
\caption{(Quasi)classical limit of the phase space states.}
\label{fig:classical_limit}
\end{figure}
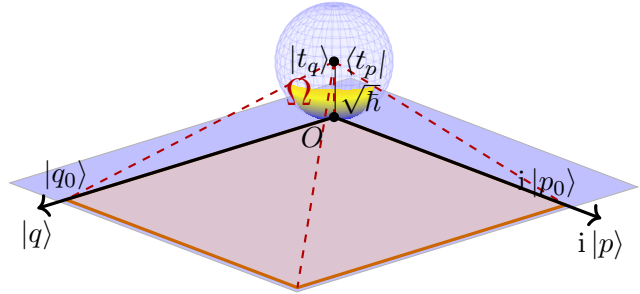
\begin{equation}
\begin{aligned}
\ket{pq}\mapsto\bigl((p_1,\ldots,p_N),(q_1,\ldots,q_N)\bigr)
&\longmapsto
\bigl((p_1,q_1),\ldots,(p_N,q_N)\bigr).
\end{aligned}
\end{equation}

Consider a hemisphere of radius $\sqrt{\hbar}$ tangent to the plane \(z=0\) at the origin.\\
The center of the sphere is therefore located at

\[
t=(0,0,\sqrt{\hbar}).
\]

Let

\[
\ket{q_0}\times\mathrm{i}\ket{p_0}
\]

be the basis where the phase space area looks non-squeezed in the tangent plane. Each point of the rectangle is projected onto the hemisphere by a central projection
from the sphere's center.
The image on the hemisphere subtends a solid angle \(\Omega\).

For an infinitesimal plane surface element \(dp\,dq\),

\begin{equation}
d\Omega
=
\frac{\cos\theta}{r^2}\,dp\,dq, \ \text{where} \ r=\sqrt{q^2+p^2+\left((\sqrt{\hbar}\right)^2},
\
\cos\theta=\frac{\sqrt{\hbar}}{r}
\end{equation}
Hence,

\begin{equation}
    d\Omega
=
\frac{\sqrt{\hbar}\,dp\,dq}
{\left(p^2+q^2+\left(\sqrt{\hbar}\right)^2\right)^{3/2}}
\end{equation}

Integrating over the rectangle gives

\begin{equation}
    \Omega(p_0,q_0;\sqrt{\hbar})
=
\int_0^{p_0}
\int_0^{q_0}
\frac{\sqrt{\hbar}\,dp\,dq}
{\left(p^2+q^2+\hbar\right)^{3/2}}=\arctan
\left(
\frac{p_0q_0}
{\sqrt{\hbar}\sqrt{\hbar+p_0^2+q_0^2}}
\right)
\end{equation}
Thus, the solid angle between the tangent vectors $\frac{\partial}{\partial q}\parallel \ket{q}$ and $\frac{\partial}{\partial p}\parallel\ket{p}$ depends slightly on the scale of the problem. And in the quantum case, we can no longer say that they are independent deviations.

\subsubsection{Quantum limit}

The exact solid angle subtended by the projected rectangle is

\begin{equation}
\Omega(p_0,q_0;\sqrt{\hbar})
=
\arctan
\left(
\frac{p_0q_0}
{\sqrt{\hbar}\sqrt{\hbar+p_0^2+q_0^2}}
\right)
\end{equation}

The natural quantity controlling the small-rectangle limit is the Kähler potential $\rho$ relation to $\hbar$:

\begin{equation}
\frac{\rho}{\hbar}=\frac{q^2+p^2}{\hbar}.
\end{equation}
Every point $(x,y)$ inside the rectangle satisfies
\begin{equation}
    x^2+y^2\le q^2+p^2,
\end{equation}

The entire rectangle remains close to the tangency point whenever

\begin{equation}\label{quantum_condition}
q^2+p^2\ll \hbar
\end{equation}

Expanding

\begin{equation}
    \frac{1}{\sqrt{1+x}}
=
1-\frac{x}{2}
+\frac{3x^2}{8}
+O(x^3),
\quad
x=\frac{q^2+p^2}{\hbar},
\end{equation}

gives

\[
\frac{pq}
{\sqrt{\hbar}\sqrt{\hbar+q^2+p^2}}
=
\frac{pq}{\hbar}
\left(
1
-
\frac{p^2+q^2}{2\hbar}
+
\frac{3(p^2+q^2)^2}{8\hbar^2}
+\cdots
\right).
\]

Since

\[
\frac{pq}{\hbar}
\le
\frac{p^2+q^2}{2\hbar}
=
\frac{\rho}{2\hbar},
\]
Using
\[
\arctan x
=
x-\frac{x^3}{3}+O(x^5),
\]

one obtains

\begin{equation}
\Omega
=
\frac{pq}{\hbar}
-
\frac{pq(p^2+q^2)}{2\hbar^2}
+\dots
\end{equation}

Hence,

\begin{equation}
\Omega
\approx
\frac{pq}{\hbar},
\qquad
p^2+q^2\ll \hbar.
\end{equation}

And inner product between states as shown in the Figure \ref{fig:quantum_limit}:
\begin{equation}\label{plane_wave}
\boxed{
    \braket{q}{p}=A\exp\left(\mathrm{i}\Omega\right)=A\exp\left(\mathrm{i}\frac{pq}{\hbar}\right)
    }
\end{equation}
Now, after the relation $\braket{p}{q}$ (eq. \ref{plane_wave}) is obtained and Hamiltonian $\hat{H}\equiv\hat{K}+\hat{U}$ is defined in \ref{quantum_potential_energy} and \ref{quantum_kinetic_energy}, the $\partial_t$ in the Schrödinger equation could be derived as described by T.W.B. Kibble in \cite{Kibble1979}. 
\subsubsection{(Quasi)classical Limit}

Now suppose $p,q\gg \sqrt{\hbar}$, introducing

\begin{equation}
   X=\frac{pq}
{\sqrt{\hbar}\sqrt{\hbar+p^2+q^2}} \Rightarrow \frac1{X^2}
=
\frac{\hbar^2}{p^2q^2}
+
\frac{\hbar}{p^2}
+
\frac{\hbar}{q^2}, \ X\to\infty
\end{equation}
Using the asymptotic expansion
\begin{equation}
    \arctan X
=
\frac{\pi}{2}
-
\frac1X
+
\frac1{3X^3}
+
\dots
\end{equation}
yields
\begin{equation}
\Omega
\approx
\frac{\pi}{2}
\end{equation}

In the limit,

\begin{equation}
    \braket{q}{p}=\exp\left(\mathrm{i}\Omega\right)=\exp\left(\mathrm{i}\frac{\pi}{2}\right)=\mathrm{i}
\end{equation}

Geometrically, an infinitely large first quadrant of the tangent plane projects
onto one quarter of the visible hemisphere, whose solid angle is
$\pi/2$, as shown in the Figure \ref{fig:classical_limit}.

\subsection{Relation to the Feynman path integral formalism}
When analyzing the Feynman path integral, we state that the principal contribution comes from slowly varying phases, while rapidly oscillating contributions cancel each other out \cite{Brown2005}. Here we first recall how this reasoning was applied in deriving the Schrödinger equation, and then, we will show how this reasoning is applied with respect to the angle between states.
\par First, in the time-sliced path integral one does not initially deal with an ordinary differentiable velocity $\dot q$, but with the finite difference
\begin{equation}
\frac{q'-q}{\varepsilon}.
\end{equation}
Accordingly, for each time slice it is not the usual Lagrangian, but the expression:
\begin{equation}
L_\varepsilon(q',q)
=
\frac{m}{2}
\left(
\frac{q'-q}{\varepsilon}
\right)^2
-
V\left(\frac{q'+q}{2}\right).
\end{equation}
The corresponding short-time propagator therefore contains the Fresnel factor
\begin{equation}
\exp\left[
\frac{im(q'-q)^2}
{2\hbar\varepsilon}
\right].
\end{equation}
As $\varepsilon\to0$, this factor becomes rapidly oscillating away from
$q'=q$, while the relevant short-time displacements scale as
\begin{equation}
q'-q\sim\sqrt{\varepsilon},
\end{equation}
formally analogous to Brownian scaling. One may therefore expand the remaining factors locally around $q'=q$, and retaining terms through $(q'-q)^2\sim\varepsilon$ gives
\begin{equation}
i\hbar\partial_t\psi
=
-\frac{\hbar^2}{2m}\partial_q^2\psi
+
V\psi,
\end{equation}
i.e. the Schrödinger equation.
\\ \\
Concerning the angle between states, let's now comment on Tobocman's \cite{Tobocman1956} way from the Schrödinger's equation to the path integral. In order to keep track of the context, let us remind it starting from eq.\ref{eq:tobocman_first} leading up to eq. \ref{eq:tobocman}.
\\
\par A one-dimensional nonrelativistic particle with Hamiltonian
\begin{equation}\label{eq:tobocman_first}
    \hat H
    =
    \frac{\hat p^2}{2m}
    +
    V(\hat q).
\end{equation}
so that the transition amplitude between the initial position
$q_0$ and the final position $q_N$ is
\begin{equation}
    K(q_N,T;q_0,0)
    =
    \left\langle q_N
    \left|
    \exp\left(
        -\frac{i}{\hbar}\hat H T
    \right)
    \right|
    q_0
    \right\rangle .
\end{equation}
We again divide the time interval $[0,T]$ into $N$ equal intervals $\epsilon=\frac{T}{N}$.
Then
\begin{equation}
    \exp\left(
        -\frac{i}{\hbar}\hat H T
    \right)
    =
    \left[
        \exp\left(
            -\frac{i}{\hbar}\hat H\epsilon
        \right)
    \right]^N .
\end{equation}

Inserting $N-1$ resolutions of the identity,
\begin{equation}
    \mathbf 1
    =
    \int_{-\infty}^{\infty}
    dq_j\,
    |q_j\rangle\langle q_j|,
\end{equation}
one obtains
\begin{equation}
\begin{split}
    K(q_N,T;q_0,0)
    =
    \left(
        \prod_{j=1}^{N-1}
        \int dq_j
    \right)
    \prod_{j=0}^{N-1}
    \left\langle
        q_{j+1}
        \left|
        \exp\left(
            -\frac{i}{\hbar}\hat H\epsilon
        \right)
        \right|
        q_j
    \right\rangle .
\end{split}
\end{equation}

For sufficiently small $\epsilon$, the Trotter product formula gives
\begin{equation}
    \exp\left[
        -\frac{i\epsilon}{\hbar}
        \left(
            \frac{\hat p^2}{2m}
            +
            V(\hat q)
        \right)
    \right]
    =
    \exp\left(
        -\frac{i\epsilon}{\hbar}
        \frac{\hat p^2}{2m}
    \right)
    \exp\left(
        -\frac{i\epsilon}{\hbar}
        V(\hat q)
    \right)
    +
    O(\epsilon^2).
\end{equation}
Hence, for a single time slice,
\begin{equation}
\begin{split}
    &
    \left\langle
        q_{j+1}
        \left|
        \exp\left(
            -\frac{i}{\hbar}\hat H\epsilon
        \right)
        \right|
        q_j
    \right\rangle
    \\
    &\qquad\simeq
    \exp\left[
        -\frac{i\epsilon}{\hbar}
        V(q_j)
    \right]
    \left\langle
        q_{j+1}
        \left|
        \exp\left(
            -\frac{i\epsilon}{\hbar}
            \frac{\hat p^2}{2m}
        \right)
        \right|
        q_j
    \right\rangle .
\end{split}
\end{equation}

We now insert the momentum-space resolution of the identity,
\begin{equation}
    \mathbf 1
    =
    \int_{-\infty}^{\infty}
    dp_j\,
    |p_j\rangle\langle p_j|,
\end{equation}
Using this resolution of the identity, the kinetic matrix element becomes
\begin{equation}
\left\langle
q_{j+1}
\left|
\exp\left(
-\frac{i\epsilon}{\hbar}
\frac{\hat p^2}{2m}
\right)
\right|
q_j
\right\rangle
\
=\int_{-\infty}^{\infty}
dp_j,
\left\langle
q_{j+1}
\left|
\exp\left(
-\frac{i\epsilon}{\hbar}
\frac{\hat p^2}{2m}
\right)
\right|
p_j
\right\rangle
\langle p_j|q_j\rangle .
\end{equation}

Since $\ket{p_j}$ is an eigenstate of the momentum operator,
$
\hat p|p_j\rangle=p_j|p_j\rangle,
$
we have
\begin{equation}
    \exp\left(
-\frac{i\epsilon}{\hbar}
\frac{\hat p^2}{2m}
\right)
|p_j\rangle
=
\exp\left(
-\frac{i\epsilon}{\hbar}
\frac{p_j^2}{2m}
\right)
|p_j\rangle .
\end{equation}

Therefore,
\begin{equation}
\left\langle
q_{j+1}
\left|
\exp\left(
-\frac{i\epsilon}{\hbar}
\frac{\hat p^2}{2m}
\right)
\right|
q_j
\right\rangle
\
=\int_{-\infty}^{\infty}
dp_j,
\exp\left(
-\frac{i\epsilon}{\hbar}
\frac{p_j^2}{2m}
\right)
\langle q_{j+1}|p_j\rangle
\langle p_j|q_j\rangle .
\end{equation}

Hence the short-time propagator can be written as
\begin{equation}\label{eq:tobocman}
\boxed{
    \left\langle
q_{j+1}
\left|
\exp\left(
-\frac{i}{\hbar}\hat H\epsilon
\right)
\right|
q_j
\right\rangle
\simeq
\int_{-\infty}^{\infty}
dp_j,
\exp\left[
-\frac{i\epsilon}{\hbar}
\left(
\frac{p_j^2}{2m}
+
V(q_j)
\right)
\right]
\langle q_{j+1}|p_j\rangle
\langle p_j|q_j\rangle .
}
\end{equation}
Now we will comment on it with respect to the angle between states. Again, the main contribution comes from slowly varying phases.
Introducing canonically rescaled phase-space coordinates:
\begin{equation}
    P=\frac{p}{\sqrt{m\omega_0}},
    \qquad
    Q=\sqrt{m\omega_0}\,q,
\end{equation}
where $\omega_0$ is a characteristic frequency scale.
\begin{equation}
    \mathcal V(Q)
    =
    \frac{1}{\omega_0}
    V\left(
        \frac{Q}{\sqrt{m\omega_0}}
    \right).
\end{equation}
\begin{equation}
    \left\langle
    Q_{j+1}
    \left|
    \exp\left(
        -\frac{i}{\hbar}\hat H\epsilon
    \right)
    \right|
    Q_j
    \right\rangle
    \simeq
    \int_{-\infty}^{\infty}
    dP_j\,
    \exp\left[
        -\frac{i\epsilon\omega_0}{\hbar}
        \left(
            \frac{P_j^2}{2}
            +
            \mathcal V(Q_j)
        \right)
    \right]
    \langle Q_{j+1}|P_j\rangle
    \langle P_j|Q_j\rangle .
\end{equation}
The principal contribution is expected to come from the region where
the corresponding phase-space frequence is small compared with $\hbar$, slow oscillations,
namely
\begin{equation}
    \frac{P_j^2}{2}
    +
    \mathcal V(Q_j)
    \ll
    \hbar,
\end{equation}

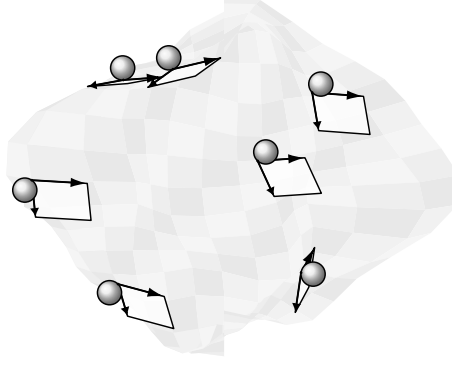
\begin{figure}[H]
\centering
\begin{tikzpicture}[
  line cap=round,
  line join=round,
  >=Latex
]


\newcommand{\ProjectPoint}[3]{%
  \pgfmathsetmacro{\PX}{#1 + 0.46*(#2)}%
  \pgfmathsetmacro{\PY}{#3 + 0.22*(#2)}%
}

%

%
%

\newcommand{\SurfacePoint}[2]{%

  \pgfmathsetmacro{\relief}{
      1
      + 0.16*sin(3*(#1) + 23)*sin(#2)^2
      + 0.11*cos(5*(#1) - 37)*sin(#2)^3
      + 0.08*sin(2*(#1) + 4*(#2) + 71)
      + 0.07*cos(7*(#1) - 3*(#2) + 19)*sin(#2)^2
      + 0.055*sin(9*(#1) + 2*(#2) - 42)*sin(#2)^3
      + 0.045*cos(4*(#1) + 7*(#2) + 11)
  }

  \pgfmathsetmacro{\RR}{
      2.35*\relief
  }

  \pgfmathsetmacro{\deformX}{
      1
      + 0.08*sin(2*(#2) + 3*(#1))
      + 0.04*cos(6*(#1) - 17)
  }

  \pgfmathsetmacro{\deformY}{
      1
      + 0.10*cos(3*(#2) - 2*(#1))
      + 0.045*sin(5*(#1) + 31)
  }

  \pgfmathsetmacro{\deformZ}{
      1
      + 0.11*sin(3*(#2) + #1 + 26)
      + 0.05*cos(5*(#2) - 4*(#1))
  }

  \pgfmathsetmacro{\SX}{
      1.02*\RR*\deformX*sin(#2)*cos(#1)
  }

  \pgfmathsetmacro{\SY}{
      0.92*\RR*\deformY*sin(#2)*sin(#1)
  }

  \pgfmathsetmacro{\SZ}{
      0.82*\RR*\deformZ*cos(#2)
  }
}

\foreach \u in {180,190,...,350}{%

  \pgfmathsetmacro{\unext}{\u+10}%

  \foreach \v in {0,10,...,170}{%

    \pgfmathsetmacro{\vnext}{\v+10}%

    \SurfacePoint{\u}{\v}
    \ProjectPoint{\SX}{\SY}{\SZ}
    \xdef\AX{\PX}
    \xdef\AY{\PY}

    \SurfacePoint{\unext}{\v}
    \ProjectPoint{\SX}{\SY}{\SZ}
    \xdef\BX{\PX}
    \xdef\BY{\PY}

    \SurfacePoint{\unext}{\vnext}
    \ProjectPoint{\SX}{\SY}{\SZ}
    \xdef\CX{\PX}
    \xdef\CY{\PY}

    \SurfacePoint{\u}{\vnext}
    \ProjectPoint{\SX}{\SY}{\SZ}
    \xdef\DX{\PX}
    \xdef\DY{\PY}

    \pgfmathtruncatemacro{\tone}{
      mod((\u-180)/10+\v/10,4)
    }

    \ifcase\tone
      \def\facetshade{gray!8}%
    \or
      \def\facetshade{gray!13}%
    \or
      \def\facetshade{gray!18}%
    \else
      \def\facetshade{gray!11}%
    \fi

    \path[
      fill=\facetshade,
      draw=none
    ]
      (\AX,\AY)
      --
      (\BX,\BY)
      --
      (\CX,\CY)
      --
      (\DX,\DY)
      --
      cycle;
  }%
}%

%

\newcommand{\TangentPatch}[2]{%

  \begingroup

  \def\hh{1.2}%
  \def\avec{0.72}%
  \def\bvec{0.58}%
  \def\spherer{0.16}%


  \SurfacePoint{#1}{#2}

  \edef\px{\SX}
  \edef\py{\SY}
  \edef\pz{\SZ}


  \pgfmathsetmacro{\up}{#1+\hh}
  \pgfmathsetmacro{\um}{#1-\hh}

  \SurfacePoint{\up}{#2}

  \edef\uxp{\SX}
  \edef\uyp{\SY}
  \edef\uzp{\SZ}

  \SurfacePoint{\um}{#2}

  \edef\uxm{\SX}
  \edef\uym{\SY}
  \edef\uzm{\SZ}

  \pgfmathsetmacro{\tx}{\uxp-\uxm}
  \pgfmathsetmacro{\ty}{\uyp-\uym}
  \pgfmathsetmacro{\tz}{\uzp-\uzm}

  \pgfmathsetmacro{\tn}{
    sqrt(\tx*\tx+\ty*\ty+\tz*\tz)
  }

  \pgfmathsetmacro{\ex}{\tx/\tn}
  \pgfmathsetmacro{\ey}{\ty/\tn}
  \pgfmathsetmacro{\ez}{\tz/\tn}


  \pgfmathsetmacro{\vp}{#2+\hh}
  \pgfmathsetmacro{\vm}{#2-\hh}

  \SurfacePoint{#1}{\vp}

  \edef\vxp{\SX}
  \edef\vyp{\SY}
  \edef\vzp{\SZ}

  \SurfacePoint{#1}{\vm}

  \edef\vxm{\SX}
  \edef\vym{\SY}
  \edef\vzm{\SZ}

  \pgfmathsetmacro{\qx}{\vxp-\vxm}
  \pgfmathsetmacro{\qy}{\vyp-\vym}
  \pgfmathsetmacro{\qz}{\vzp-\vzm}

  \pgfmathsetmacro{\edotq}{
    \ex*\qx+\ey*\qy+\ez*\qz
  }

  \pgfmathsetmacro{\rx}{
    \qx-\edotq*\ex
  }

  \pgfmathsetmacro{\ry}{
    \qy-\edotq*\ey
  }

  \pgfmathsetmacro{\rz}{
    \qz-\edotq*\ez
  }

  \pgfmathsetmacro{\rn}{
    sqrt(\rx*\rx+\ry*\ry+\rz*\rz)
  }

  \pgfmathsetmacro{\fx}{\rx/\rn}
  \pgfmathsetmacro{\fy}{\ry/\rn}
  \pgfmathsetmacro{\fz}{\rz/\rn}


  \pgfmathsetmacro{\nx}{
    \ey*\fz-\ez*\fy
  }

  \pgfmathsetmacro{\ny}{
    \ez*\fx-\ex*\fz
  }

  \pgfmathsetmacro{\nz}{
    \ex*\fy-\ey*\fx
  }

  \pgfmathsetmacro{\outdot}{
    \nx*\px+\ny*\py+\nz*\pz
  }

  \pgfmathsetmacro{\sgn}{
    ifthenelse(\outdot<0,-1,1)
  }

  \pgfmathsetmacro{\nx}{\sgn*\nx}
  \pgfmathsetmacro{\ny}{\sgn*\ny}
  \pgfmathsetmacro{\nz}{\sgn*\nz}


  \pgfmathsetmacro{\qax}{
    \px+\avec*\ex
  }

  \pgfmathsetmacro{\qay}{
    \py+\avec*\ey
  }

  \pgfmathsetmacro{\qaz}{
    \pz+\avec*\ez
  }

  \pgfmathsetmacro{\qbx}{
    \px+\bvec*\fx
  }

  \pgfmathsetmacro{\qby}{
    \py+\bvec*\fy
  }

  \pgfmathsetmacro{\qbz}{
    \pz+\bvec*\fz
  }

  \pgfmathsetmacro{\qcx}{
    \px+\avec*\ex+\bvec*\fx
  }

  \pgfmathsetmacro{\qcy}{
    \py+\avec*\ey+\bvec*\fy
  }

  \pgfmathsetmacro{\qcz}{
    \pz+\avec*\ez+\bvec*\fz
  }


  \ProjectPoint{\px}{\py}{\pz}

  \edef\PPX{\PX}
  \edef\PPY{\PY}

  \ProjectPoint{\qax}{\qay}{\qaz}

  \edef\PAX{\PX}
  \edef\PAY{\PY}

  \ProjectPoint{\qbx}{\qby}{\qbz}

  \edef\PBX{\PX}
  \edef\PBY{\PY}

  \ProjectPoint{\qcx}{\qcy}{\qcz}

  \edef\PCX{\PX}
  \edef\PCY{\PY}

  %

  \ProjectPoint{\nx}{\ny}{\nz}

  \pgfmathsetmacro{\nnx}{\PX}
  \pgfmathsetmacro{\nny}{\PY}

  \pgfmathsetmacro{\nnorm}{
    sqrt(\nnx*\nnx+\nny*\nny)
  }

  \pgfmathsetmacro{\cx}{
    \PPX+\spherer*\nnx/\nnorm
  }

  \pgfmathsetmacro{\cy}{
    \PPY+\spherer*\nny/\nnorm
  }


  \path[
    fill=white,
    fill opacity=0.82,
    draw=black,
    line width=0.55pt
  ]
    (\PPX,\PPY)
    --
    (\PAX,\PAY)
    --
    (\PCX,\PCY)
    --
    (\PBX,\PBY)
    --
    cycle;


  \draw[
    -{Latex[length=2.2mm,width=1.25mm]},
    line width=0.72pt
  ]
    (\PPX,\PPY)
    --
    (\PAX,\PAY);

  \draw[
    -{Latex[length=1.2mm,width=1.25mm]},
    line width=0.72pt
  ]
    (\PPX,\PPY)
    --
    (\PBX,\PBY);


  \shade[
    ball color=gray!18,
    draw=black,
    line width=0.55pt
  ]
    (\cx,\cy)
    circle[radius=\spherer];

  \fill
    (\PPX,\PPY)
    circle[radius=0.018];

  \endgroup
}


\TangentPatch{245}{40}
\TangentPatch{240}{80}
\TangentPatch{274}{42}
\TangentPatch{307}{70}
\TangentPatch{337}{48}
\TangentPatch{250}{125}
\TangentPatch{316}{121}

\end{tikzpicture}

\caption{Main contributors (slow oscillations) to the path integral in case of quadratic potentials are building the slowly oscillating phase space volume flow.}
\label{fig:path_integral}

\end{figure}

A particularly important class is formed by classical Hamiltonian
functions that are at most quadratic in
the canonical variables. This includes the harmonic oscillator, the
free particle, a particle in a linear potential, and more generally
any Hamiltonian of the form
\begin{equation}
    \mathcal H(Q,P)
    =
    aP^2+bPQ+cQ^2+dP+eQ+f.
\end{equation}
Here $Q$ and $P$ are commuting phase-space coordinates. The mixed term is represented by
$
    bPQ
    \longmapsto
    \frac{b}{2}
    \left(
        \hat P\hat Q+\hat Q\hat P
    \right).
$
Affine linear canonical transformations reduce such Hamiltonians to
quadratic normal forms in many cases.
\\ Then
condition that the corresponding phase-space oscillation be slow acquires a condition:
\begin{equation}
    P^2+Q^2\ll\hbar.
\end{equation}
That is, now we present the condition for the Tobocman's calculation \cite{Tobocman1956}, according to the \ref{quantum_condition}:
\begin{equation}
    \langle Q|P\rangle
    =\frac{1}{\sqrt{2\pi\hbar}}\exp\left(\mathrm{i}\arctan
\left(
\frac{PQ}
{\sqrt{\hbar}\sqrt{\hbar+P^2+Q^2}}
\right)\right) \ \longrightarrow \
    \frac{1}{\sqrt{2\pi\hbar}}
    \exp\left(
        \frac{i}{\hbar}PQ
    \right), \ P^2+Q^2\ll\hbar.
\end{equation}
It follows that
\begin{equation}
    \left\langle
    Q_{j+1}
    \left|
    \exp\left(
        -\frac{i}{\hbar}\hat H\epsilon
    \right)
    \right|
    Q_j
    \right\rangle
    \simeq
    \int_{-\infty}^{\infty}
    \frac{dP_j}{2\pi\hbar}\,
    \exp\left\{
        \frac{i}{\hbar}
        \left[
            P_j(Q_{j+1}-Q_j)
            -
            \epsilon\omega_0
            \left(
                \frac{P_j^2}{2}
                +
                \mathcal V(Q_j)
            \right)
        \right]
    \right\}.
\end{equation}
And, Fig.\ref{fig:path_integral}:
\begin{equation}
    K(Q_N,T;Q_0,0)
    =
    \lim_{N\to\infty}
    \int
    \prod_{j=1}^{N-1}dQ_j
    \prod_{j=0}^{N-1}
    \frac{dP_j}{2\pi\hbar}
    \times
    \exp\left\{
        \frac{i}{\hbar}
        \sum_{j=0}^{N-1}
        \left[
            P_j(Q_{j+1}-Q_j)
            -
            \epsilon\omega_0
            \left(
                \frac{P_j^2}{2}
                +
                \mathcal V(Q_j)
            \right)
        \right]
    \right\}.
\end{equation}

\section{Comment on the Born rule.}
\subsection{Free semigroup evolution}
In this section, we assume that every measurement implicitly contains the measurement of the $\ket{t}-$approximation, but not the parameter $t$. In other words, in addition to a measurement operator, we stress on the ket to be measured. We also need something other ("detector") than the free pure system itself, in order to interact with the system/measure the system state. We will remain in the laboratory reference frame from the first measurement to the $n$th measurement.\\
We consider $\ket{t}$ in def. \ref{t_definition} as a state already, thus we omit the field $\abs{\Psi}$ in the Schrödinger equation:
\begin{equation}
    \mathrm{i}\hbar\frac{\dd \ket{t}}{\dd t}=\hat{H}\ket{t}
\end{equation}
And then if $\hat{H}$ does not depend on $t$:
\begin{equation}
    \ket{t}=\exp{-\frac{\mathrm{i}}{\hbar}\hat{H}t}\ket{0}
\end{equation}
This transformation is unitary; therefore, it preserves the time length $\Rightarrow$ no new time blocks appear.
The next reasoning is based on the example: when we take two different $\hat{H_L}$ and $\hat{H_R}$, they will separate two equally prepared states $\ket{t_0}$:
\begin{equation}
        \ket{t_L}=\exp{-\frac{\mathrm{i}}{\hbar}\hat{H_L}t}\ket{t_0}; \qquad \ket{t_R}=\exp{-\frac{\mathrm{i}}{\hbar}\hat{H_R}t}\ket{t_0}; 
\end{equation}
It is possible that $\ket{t_i}\perp\ket{t_j}$ even if the parameter $t$ is the same.
\subsection{Double-slit experiment}
 We compare the measurements of a single state and the one of two superposition states, for example, in the double-slit experiment. We will not actually be examining the interference pattern on the screen; instead, we will look at the Born rule in a case of two states.
\begin{equation}
    \ket{q} \ -\ \text{particle position state,} \ \ket{p} \ -\ \text{state of the detector.} 
\end{equation}
\subsubsection{Detector is far away from the slit}
When detector is far away from the slit, we measure the time-entangled state of the single particle $\ket{q}$ and detector $\ket{p}$ states. When a free particle is hitting free detector we have the entangling evolution:
\begin{equation}
    \braket{q}{t_0}\otimes\braket{p}{t_0}\mapsto\braket{qp}{t}
\end{equation}
The $\mapsto$ is equal to turning on the $H_\text{int}$ at $t=t_0:$
\begin{multline}
    \ket{\psi(t)}
=
\mathcal{T}\exp\!\left[-\frac{i}{\hbar}\int_{0}^{t}\!\left(\hat H_{\mathrm{free}}+\Theta(\tau-t_0)\hat H_{\mathrm{int}}\right)d\tau\right]\ket{\psi(0)}\\
=
\exp\!\left[-\frac{i}{\hbar}(\hat H_{\mathrm{free}}+\hat H_{\mathrm{int}})(t-t_0)\right]
\exp\!\left[-\frac{i}{\hbar}\hat H_{\mathrm{free}}t_0\right]\ket{\psi(0)}.
\end{multline}
That is
\begin{equation}
    \braket{q}{\exp{-\frac{i}{\hbar}\hat{H}_\text{free}t_0}|0}\otimes\braket{p}{\exp{-\frac{i}{\hbar}\hat{H}_\text{free}t_0}|0}\longmapsto\braket{qp}{\exp{-\frac{i}{\hbar}\hat{H}_\text{interaction}t}|t_{00}}
\end{equation}
Which is tantamount to:
\begin{equation}
    \ket{\exp{-\frac{i}{\hbar}\hat{H}_qt_0}|0}\otimes\ket{\exp{-\frac{i}{\hbar}\hat{H}_pt_0}|0}\equiv\ket{t_{00}}\longmapsto\ket{\exp{-\frac{i}{\hbar}\hat{H}_{qp}t}|t_{00}}=\ket{t_{qp}}
\end{equation}

\subsubsection{Detector is at the double slit}
We remind that transformations here are unitary; therefore, it preserves the time length $\Rightarrow$ no new time blocks appear: $\|\ket{t}\|^2=\text{const}$.
In details, after any evolution (free or with interaction/entangling) the unitary $U(t)$ gives:
\begin{equation}\label{eq:unitarity}
    \|u(0)\otimes v(0)\|=\|U_u(t)u(0)\otimes U_v(t)v(0)\|\equiv\|\omega(0)\|=\|u(t)\otimes v(t)\|=\|U_{uv}(\tau-t)\left[u(t)\otimes v(t)\right]\|=\|w(\tau)\|
\end{equation}
\par When detector is sensitive to the individual left $\ket{q_L}$ and right $\ket{q_R}$ slits of the double slit then for the initially prepared state $C_L\bra{q_L}+C_R\bra{q_R}$ we have:
\begin{equation}
    \braket{q}{t_0}\otimes\braket{p}{t_0}=\big[C_L\bra{q_L}+C_R\bra{q_R}\big]\ket{t_0}\otimes\braket{p}{t_0}
\end{equation}
$$
    C_L\ket{\exp{-\frac{i}{\hbar}\hat{H}_{q_L}t_0}|0}\otimes\ket{\exp{-\frac{i}{\hbar}\hat{H}_pt_0}|0}\longmapsto C_L\ket{\exp{-\frac{i}{\hbar}\hat{H}_{q_Lp}t}|t_{00}}=C_L\ket{t_L}
$$
\begin{equation}
    C_R\ket{\exp{-\frac{i}{\hbar}\hat{H}_{q_R}t_0}|0}\otimes\ket{\exp{-\frac{i}{\hbar}\hat{H}_pt_0}|0}\longmapsto C_R\ket{\exp{-\frac{i}{\hbar}\hat{H}_{q_Rp}t}|t_{00}}=C_R\ket{t_R}
\end{equation}

The $1\cdot\ket{t_L}$ and $1\cdot\ket{t_R}$ above represent different experimental sets, when one of two slits has to be closed.\\
If we open both slits, the final states $\ket{L}$ and $\ket{R}$ still should not be possible to measure simultaneously with the detector at the double slit. And multiple runs of this same experiment should lead us to: \\

\begin{equation}\label{constant_times_semigroup}
\begin{pmatrix}
t_{{00}_1}\\
t_{{00}_2}\\
t_{{00}_3}\\
t_{{00}_4}\\
t_{{00}_5}\\
t_{{00}_6}\\
\vdots\\
t_{{00}_\infty}
\end{pmatrix}
\quad\longmapsto\quad
\bra{C_L}
\begin{pmatrix}
t_{{L}_1}\\
t_{{L}_2}\\
t_{{L}_3}\\
\vdots\\
t_{{L}_N}
\end{pmatrix}
+\bra{C_R}
\begin{pmatrix}
t_{R_1}\\
t_{R_2}\\
t_{R_3}\\
\vdots\\
t_{{R}_M}
\end{pmatrix}
\quad =\quad
\underbrace{
\begin{pmatrix}
t_{{L}_1}\\
\mathbf{0}\\
t_{{L}_3}\\
\mathbf{0}\\
\mathbf{0}\\
\vdots\\
t_{{L}_{\varepsilon_N}}
\end{pmatrix}
}_{\braket{C_L}{t_L}}
+
\underbrace{
\begin{pmatrix}
\mathbf{0}\\
t_{R_2}\\
\mathbf{0}\\
t_{R_4}\\
t_{R_5}\\
\vdots\\
\mathbf{0}
\end{pmatrix}
}_{\braket{C_R}{t_R}}
\end{equation}
The problem of multiplying a one-dimensional constant by a free semigroup element arose. To define such an operation in \ref{constant_times_semigroup}, we recall that the greater the length of the free semigroup element, the more non-zero letters it contains. We also have to define it in a vector space. The constant will alter the length of the free semigroup element, now a vector, in $\mathbb{C}[\ket{t}]$, thereby changing the proportion of non-zero states within it. Further we describe how the constant is, as it were, split into the letters of free semigroup element.\\
In other words, we initially think of pure numbers as some abstract self-distributors.\\ We also remember  (eq.\ref{eq:unitarity}):

\begin{equation}
    \|\braket{q}{t_0}\otimes\braket{p}{t_0}\|\equiv\|\ket{t_{00}}\|\equiv1=\|C_L\ket{t_L}+C_R\ket{t_R}\| 
\end{equation}

\subsubsection{Constant acts on a free $\ket{t}$ semigroup element}
Consider a finite word
\begin{equation}
    \ket{t_L}=\ket{t_1}+\ket{t_2}+\ket{t_3}+...+\ket{t_N},
\end{equation}
with the Pythagorean length
\begin{equation}
    \|\ket{t_L}\|^2
=
\sum_{k=1}^{N}\|\ket{t_k}\|^2.
\end{equation}

In particular, if every generator has unit length,
$
\|\ket{t_k}\|=1,
$
then
$
\|\ket{t_L}\|^2=N
$, 

Let
\begin{equation}
    C_L=\abs{C_L}e^{i\theta}\in\mathbb{C},
\qquad
0\leq \abs{C_L}<1,
\end{equation}
Let
\begin{equation}
    \xi_1,\ldots,\xi_N
\end{equation}
be independent Bernoulli random variables with probability $\mathrm{p}:$
\begin{equation}
    \mathbb{P}(\xi_k=1)=\mathrm{p},
\qquad
\mathbb{P}(\xi_k=0)=1-\mathrm{p}.
\end{equation}

Analogously to the $\braket{q}{t}$ in eq. \ref{def:position_with_0}, we are trying to get the trajectory of the "repeated" experiments, but out of only one number. Thus, we think of pure numbers as some abstract self-distributors.\\
So now the trajectory $\braket{q}{t}$ has the form:
\begin{equation}
\braket{C_L}{t_L}=\bra{C_L}\odot
\begin{pmatrix}
t_1\\
t_2\\
\vdots\\
t_N
\end{pmatrix}
=e^{i\theta}
\begin{pmatrix}
\braket{\xi_1}{t_1}\oplus
\braket{\xi_2}{t_2}\oplus
...\oplus
\braket{\xi_N}{t_N}
\end{pmatrix}
\end{equation}
The modulus \(|C_L|\) will self-distribute inside the free semigroup element, and the phase $\theta$ distributes the free semigroup elements at the circle. \\

\begin{equation}
   =e^{i\theta}\left[
\ket{t_{i_1}}+\ket{t_{i_2}}+\ket{t_{i_3}}+...+\ket{t_{i_{\varepsilon_N}}}\right]\equiv\ket{C_Lt_L}
\end{equation}
where
\begin{equation}
    1\leq i_1<i_2<\cdots<i_{\varepsilon_N}\leq N
\end{equation}
and the number of surviving substates (or augmentation (eq.\ref{def:augmentation}), $\varepsilon|e^{i\theta}=e^{-i\theta}|e^{i\theta}$):
\begin{equation}
   \braket{\varepsilon}{C_Lt_L}\equiv \sum_{k=1}^{N}\xi_k\equiv \varepsilon_N
\end{equation}

so the trajectory now is the thinning of \(\ket{t_L}\), since $\abs{C_L}\le1$, and we get the subword.\\
\subsubsection{The Born rule}

For unit-length generators,
\begin{equation}
    \|\braket{C_L}{t_L}\|^2
=
\varepsilon_N
\ \Rightarrow \
   \frac{
\|\braket{C_L}{t_L}\|^2
}{
\|\ket{t_L}\|^2
}
=
\frac{\varepsilon_N}{N} 
\end{equation}

\begin{equation}\mathbb E\!\left[\frac{
\|\braket{C_L}{t_L}\|^2
}{
\|\ket{t_L}\|^2
}\right]=
\mathbb E\!\left[\frac{\varepsilon_N}{N}\right]
=
\mathbb E\!\left[
\frac{1}{N}\sum_{k=1}^{N}\xi_k 
\right]
=
\frac{1}{N}\sum_{k=1}^{N}\mathbb E[\xi_k] 
=
\frac{1}{N}\sum_{k=1}^{N}\mathrm{p}
=
\mathrm{p}.
\end{equation}

From the other side, we consider \(\mathbb{C}[\ket{t
}]\) as Hilbert space and avoid probability-averaging:
\begin{equation}
\mathbb E\!\left[\frac{
\|\braket{C_L}{t_L})\|^2
}{
\|\ket{t_L}\|^2
}\right]=\mathbb E\!\left[\frac{
\abs{C_L}^2\|\ket{t_L}\|^2
}{
\|\ket{t_L}\|^2
}\right]=\frac{
\abs{C_L}^2\|\ket{t_L}\|^2
}{
\|\ket{t_L}\|^2
}=\abs{C_L}^2
\end{equation}
Thus,
\begin{equation}
    \mathrm{p}=\abs{C_L}^2
\end{equation}
And due to unitarity:
\begin{equation}
    \abs{C_R}^2=1-\abs{C_L}^2=1-\mathrm{p}
\end{equation}
For non-unit generator lengths, the same relation should remain valid.\\
And for the $N\to\infty$ by the strong law of large numbers,
\begin{equation}
    \frac{\varepsilon_N}{N}
\longrightarrow
\mathrm{p}=|C_L|^2
\qquad
\text{almost surely as }N\to\infty.
\end{equation}
is the Born rule.
\subsection{Mixed state vector}
Now we can define natural vectors for the density tensor, the mixed state vector as:
\begin{equation}
    \ket{\text{mixed state}}=\ket{q_L}\braket{q_L}{t_L}+\ket{q_R}\braket{q_R}{t_R}
\end{equation}
And superposition will be then:
\begin{equation}
    \ket{\text{superposition state}}=\ket{q_L}\braket{q_L}{t}+\ket{q_R}\braket{q_R}{t}
\end{equation}
The probabilities and complex constants are hidden into the corresponding kets.
\section{Acknowledgements}
I thank  Evgeny Goryachko for his algebra lectures I had the luck to contemplate, for the detailed account of the mathematics whose existence I could have barely imagined. I thank Evgeny Goryachko for the new translation of the mathematical sense for physicists and for the inspiration.\\
I thank Vladimir Gorkov for the stress on the uncharted abstract nature of the state vector, the abstract exhortation and for the remarks on the draft of this work.

\bibliographystyle{apsrev4-2}
\bibliography{main}

@article{Moyal1949,
  title = {Quantum mechanics as a statistical theory},
  volume = {45},
  ISSN = {1469-8064},
  url = {http://dx.doi.org/10.1017/S0305004100000487},
  DOI = {10.1017/s0305004100000487},
  number = {1},
  journal = {Mathematical Proceedings of the Cambridge Philosophical Society},
  publisher = {Cambridge University Press (CUP)},
  author = {Moyal,  J. E.},
  year = {1949},
  month = Jan,
  pages = {99–124}
}

@article{Groenewold1946,
  title = {On the principles of elementary quantum mechanics},
  volume = {12},
  ISSN = {0031-8914},
  url = {http://dx.doi.org/10.1016/S0031-8914(46)80059-4},
  DOI = {10.1016/s0031-8914(46)80059-4},
  number = {7},
  journal = {Physica},
  publisher = {Elsevier BV},
  author = {Groenewold,  H.J.},
  year = {1946},
  month = Oct,
  pages = {405–460}
}

@article{Wigner1932,
  title = {On the Quantum Correction For Thermodynamic Equilibrium},
  volume = {40},
  ISSN = {0031-899X},
  url = {http://dx.doi.org/10.1103/PhysRev.40.749},
  DOI = {10.1103/physrev.40.749},
  number = {5},
  journal = {Physical Review},
  publisher = {American Physical Society (APS)},
  author = {Wigner,  E.},
  year = {1932},
  month = June,
  pages = {749–759}
}

@article{Weyl1927,
  title = {Quantenmechanik und Gruppentheorie},
  volume = {46},
  ISSN = {1434-601X},
  url = {http://dx.doi.org/10.1007/BF02055756},
  DOI = {10.1007/bf02055756},
  number = {1-2},
  journal = {Zeitschrift f\"{u}r Physik},
  publisher = {Springer Science and Business Media LLC},
  author = {Weyl,  H.},
  year = {1927},
  month = Nov,
  pages = {1–46}
}

@article{Berezin1974,
  title = {QUANTIZATION},
  volume = {8},
  ISSN = {0025-5726},
  url = {http://dx.doi.org/10.1070/IM1974v008n05ABEH002140},
  DOI = {10.1070/im1974v008n05abeh002140},
  number = {5},
  journal = {Mathematics of the USSR-Izvestiya},
  publisher = {Steklov Mathematical Institute},
  author = {Berezin,  F A},
  year = {1974},
  month = Oct,
  pages = {1109–1165}
}

@article{deGosson2012,
  title = {Quantum Blobs},
  volume = {43},
  ISSN = {1572-9516},
  url = {http://dx.doi.org/10.1007/s10701-012-9636-x},
  DOI = {10.1007/s10701-012-9636-x},
  number = {4},
  journal = {Foundations of Physics},
  publisher = {Springer Science and Business Media LLC},
  author = {de Gosson,  Maurice A.},
  year = {2012},
  month = Feb,
  pages = {440–457}
}

@article{Dirac1939,
  title = {A new notation for quantum mechanics},
  volume = {35},
  ISSN = {1469-8064},
  url = {http://dx.doi.org/10.1017/S0305004100021162},
  DOI = {10.1017/s0305004100021162},
  number = {3},
  journal = {Mathematical Proceedings of the Cambridge Philosophical Society},
  publisher = {Cambridge University Press (CUP)},
  author = {Dirac,  P. A. M.},
  year = {1939},
  month = July,
  pages = {416–418}
}

@book{Mingo2017,
  title = {Free Probability and Random Matrices},
  ISBN = {9781493969425},
  ISSN = {2194-3079},
  url = {http://dx.doi.org/10.1007/978-1-4939-6942-5},
  DOI = {10.1007/978-1-4939-6942-5},
  journal = {Fields Institute Monographs},
  publisher = {Springer New York},
  author = {Mingo,  James A. and Speicher,  Roland},
  year = {2017}
}

@article{Tobocman1956,
  title = {Transition amplitudes as sums over histories},
  volume = {3},
  ISSN = {1827-6121},
  url = {http://dx.doi.org/10.1007/BF02785004},
  DOI = {10.1007/bf02785004},
  number = {6},
  journal = {Il Nuovo Cimento},
  publisher = {Springer Science and Business Media LLC},
  author = {Tobocman,  W.},
  year = {1956},
  month = June,
  pages = {1213–1229}
}

@book{Brown2005,
  title = {Feynman’s Thesis — A New Approach to Quantum Theory},
  ISBN = {9789812567635},
  url = {http://dx.doi.org/10.1142/5852},
  DOI = {10.1142/5852},
  publisher = {WORLD SCIENTIFIC},
  author = {Brown,  Laurie M},
  year = {2005},
  month = Aug 
}

@article{Kibble1979,
  title = {Geometrization of quantum mechanics},
  volume = {65},
  ISSN = {1432-0916},
  url = {http://dx.doi.org/10.1007/BF01225149},
  DOI = {10.1007/bf01225149},
  number = {2},
  journal = {Communications in Mathematical Physics},
  publisher = {Springer Science and Business Media LLC},
  author = {Kibble,  T. W. B.},
  year = {1979},
  month = June,
  pages = {189–201}
}

@book{CannasDaSilva2001,
  author    = {Cannas da Silva, Ana},
  title     = {Lectures on Symplectic Geometry},
  year      = {2001},
  }

\end{document}